\documentclass[12pt]{article}
\usepackage{amsmath,amssymb,epsfig}
\makeatletter \@addtoreset{equation}{section} \makeatother
\renewcommand{\theequation}{\thesection.\arabic{equation}}
\newcommand{\ba}{\begin{array}}
\newcommand{\ea}{\end{array}}
\newcommand{\beq}{\begin{equation}}
\newcommand{\eeq}{\end{equation}}
\newcommand{\bea}{\begin{eqnarray}}
\newcommand{\eea}{\end{eqnarray}}
\def\bce{\begin{center}}
\def\ece{\end{center}}

\def\nonu{\nonumber}

\def\pa{\partial}

\def\be{\beta}

\newcommand{\tr}{\mbox{Tr}}

\def\eps6{{\displaystyle \mathop{\epsilon}^{6}}{}}

\def\nab6{{\displaystyle \mathop{\nabla}^{6}}{}}
\def\0{{\sst{(0)}}}
\def\1{{\sst{(1)}}}
\def\2{{\sst{(2)}}}
\def\3{{\sst{(3)}}}
\def\4{{\sst{(4)}}}
\def\5{{\sst{(5)}}}
\def\6{{\sst{(6)}}}
\def\7{{\sst{(7)}}}
\def\8{{\sst{(8)}}}

\def\ba{\begin{array}}
\def\ea{\end{array}}
\def\beq{\begin{equation}}
\def\eeq{\end{equation}}
\def\be{\begin{equation}}
\def\ee{\end{equation}}

\def\tr{\mathop{\rm tr}}

\def\eps{\epsilon}

\def\ba{\begin{array}}
\def\ea{\end{array}}
\def\beq{\begin{equation}}
\def\eeq{\end{equation}}
\def\be{\begin{equation}}
\def\ee{\end{equation}}

\def\tr{\mathop{\rm tr}}

\def\eps{\epsilon}

\newcommand{\bean}{\begin{eqnarray*}}
\newcommand{\eean}{\end{eqnarray*}}

\begin{document}
\thispagestyle{empty} \addtocounter{page}{-1}
\begin{flushright}
%KIAS-P06003 \\
%CALT-68-nnnn \\
%{\tt hep-th/0701145}\\
\end{flushright}

\vspace*{1.3cm}

\centerline{ \Large \bf Other Meta-Stable Brane Configuration }
\vspace{.3cm} 
\centerline{ \Large \bf  by Adding an Orientifold 6-Plane to Giveon-Kutasov} 
\vspace*{1.5cm}
\centerline{{\bf Changhyun Ahn} 
%and {\bf Yutaka Ookouchi $^{2}$}
} 
\vspace*{1.0cm} 
\centerline{\it 
Department of Physics, Kyungpook National University, Taegu
702-701, Korea} 
%\centerline{\it $^{2}$ California Institute of 
%Technology, Pasadena, CA91125, USA }
\vspace*{0.8cm} 
\centerline{\tt ahn@knu.ac.kr} 
%\qquad
%yutaka@caltech.edu} 
\vskip2cm

\centerline{\bf Abstract}
\vspace*{0.5cm}

Giveon and Kutasov have found the type IIA
intersecting nonsupersymmetric meta-stable brane
configuration where the electric gauge theory superpotential has 
a quartic term as well as the mass term for quarks.
In this paper, by adding the orientifold 6-plane to this brane
configuration,
we describe the brane configuration 
corresponding to the meta-stable nonsupersymmetric 
vacua of the supersymmetric unitary gauge theory with symmetric flavor
as well as fundamental flavors.  

\baselineskip=18pt
\newpage
\renewcommand{\theequation}
{\arabic{section}\mbox{.}\arabic{equation}}

%%%%%%%%%%%%%%%%%%%%%%%%
\section{Introduction}
%%%%%%%%%%%%%%%%%%%%%%%%

The dynamical supersymmetry breaking in meta-stable vacua \cite{ISS} 
occurs 
in the standard ${\cal N}=1$ SQCD with massive fundamental 
flavors.
For recent developments
on supersymmetry breaking in various contexts, it is useful to see 
the review paper \cite{IS}.
The extra mass term
for quarks in the magnetic superpotential leads to the fact that
some of the F-term equations cannot  be satisfied and then the
supersymmetry is broken.
The meta-stable brane realizations of type IIA string theory
corresponding to ${\cal N}=1$ SQCD with massive  fundamental flavors 
have been found in \cite{OO1,FGU,BGHSS}. For 
the brane dynamics and supersymmetric gauge theory in the context of
supersymmetric ground states, 
the review paper \cite{GK98} is also very useful. 
 
Recently, Giveon and Kutasov \cite{GK0710-1} have found 
the type IIA intersecting nonsupersymmetric meta-stable 
brane configuration consisting of 
one NS5'-brane, one NS5-brane, D4-branes and ``rotated'' D6-branes.
They also have 
studied the corresponding gauge theory in \cite{GK0710} by using only
field theory analysis
and described the supersymmetric and nonsupersymmetric meta-stable 
ground states.
The meta-stable vacua of \cite{GK0710} occur in some
region of parameter space when some of the flavor 
D4-branes, in \cite{GK0710-1}, 
connecting the NS5'-brane and ``rotated'' D6-branes approach to 
the NS5-brane. 
The mass term for the meson field in the magnetic superpotential 
corresponds to the rotation of D6-branes 
along the (45)-(89) directions and can be interpreted as a ``quartic''
term for quarks in the corresponding 
${\cal N}=1$ electric SQCD with massive
fundamental flavors. Note that classically there exist only
supersymmetric ground states and no meta-stable ground states. 

Now
one can add an orientifold 6-plane(O6-plane), located at the NS5-brane, 
into the brane configuration of 
\cite{GK0710-1} together with an extra NS5'-brane and the mirrors for
D4-branes and ``rotated'' D6-branes. 
Then 
the type IIA brane configuration consists of 
two NS5'-branes, one NS5-brane, 
D4-branes, ``rotated'' D6-branes and an O6-plane.
It turns out that the meta-stable nonsupersymmetric 
brane configuration can be determined from the
brane configuration of \cite{GK0710-1} with the presence of O6-plane. 
We'll see how the corresponding supersymmetric gauge theory, which is
a standard ${\cal N}=1$ SQCD with massive flavors by insertion of an
extra matter content, occurs.

In this paper, we study ${\cal N}=1$ $SU(N_c)$ gauge theory with 
a symmetric flavor $S$, a conjugate symmetric flavor $\widetilde{S}$
and 
$N_f$ fundamental
flavors $Q$ and $\widetilde{Q}$ 
in the context of dynamical supersymmetric breaking vacua.
The corresponding supersymmetric brane configuration in type IIA
string theory was found sometime ago.
Now we deform this theory by  adding  both the mass
term and ``quartic'' term for quarks 
in the fundamental representation of the gauge group \cite{GK0710-1}. 
Then we turn to the
dual magnetic  gauge theory by standard brane motion \cite{Ahn07}. 
The dual
magnetic theory giving rise to the meta-stable vacua 
is described by ${\cal N}=1$
$SU(2N_f-N_c)$ gauge theory with dual matter contents. The
magnetic superpotential consists of  a linear term and quadratic term in
a meson field plus the coupling term between the meson and
dual matters. By analyzing this superpotential, along the line of 
\cite{GK0710-1,GK0710}, we present the behaviors of gauge theory
description and string theory description.

In section 2, we consider the type IIA brane configuration corresponding
to the electric theory based on the ${\cal N}=1$ $SU(N_c)$ gauge theory
with above matter contents. 
The rotation of D6-branes is crucial.
In section 3, we construct the Seiberg dual magnetic theory which is 
${\cal N}=1$ $SU(2N_f-N_c)$ gauge theory with corresponding dual
matters by brane motion. The rotation of D6-branes is encoded in the
mass term for the meson field in the superpotential.
In section 4, we describe the nonsupersymmetric meta-stable
minimum both in the gauge theory analysis and string theory analysis
and present 
the corresponding intersecting brane configuration of type IIA string
theory, along the lines of \cite{Ahn06}-\cite{Ahn07-10}.
In section 5, we make our summary of this paper and comment on the
future works \footnote{There is an alternative approach type IIB
geometrical engineering by recent work \cite{TW0711} which is a
T-dual to \cite{GK0710-1}.}.

%%%%%%%%%%%%%%%%%%%%%%%%%%%%%%%%%%%%%%%%%%%%%%%%%%%%%%%%%%%%%%%%%%%%%%
%%%%%%%%%%%%%%%%%%%%%%%%%%%%%%%%%%%%%%%%%%%%%%%%%%%%%%%%%%%%%%%%%%%%%%
\section{The ${\cal N}=1$ supersymmetric electric brane configuration}
%%%%%%%%%%%%%%%%%%%%%%%%%%%%%%%%%%%%%%%%%%%%%%%%%%%%%%%%%%%%%%%%%%%%%%
%%%%%%%%%%%%%%%%%%%%%%%%%%%%%%%%%%%%%%%%%%%%%%%%%%%%%%%%%%%%%%%%%%%%%%

The type IIA supersymmetric electric
brane configuration \cite{LL,LLL,Ahn07} corresponding to 
${\cal N}=1$ $SU(N_c)$ gauge theory  with 
a symmetric flavor $S$, a conjugate symmetric flavor $\widetilde{S}$ and 
$N_f$ fundamental flavors $Q, \widetilde{Q}$ \cite{ILS}
can be described as follows: one middle NS5-brane(012345), two
NS5'-branes(012389) denoted by $NS5_L'$-brane and $NS5_R'$-brane 
respectively, 
$N_c$
D4-branes(01236)
between them, $2N_f$ D6-branes(0123789) and an
orientifold 6 plane(0123789) of positive RR charge. 
The transverse coordinates $(x^4, x^5, x^6)$ transform as $(-x^4, -x^5,
-x^6)$ under the orientifold 6-plane(O6-plane) action. 
We introduce two complex coordinates \cite{GK98}
\bea
v \equiv x^4 + i x^5, \qquad w \equiv x^8 + i x^9.
\nonu
\eea
Then the origin of
the coordinates $(x^6, v, w)$ is located at the intersection of
NS5-brane and O6-plane.
The left $NS5_L'$-brane is located at the left hand side of O6-plane
while the right $NS5_R'$-brane is located at the right hand side of 
O6-plane. The $N_c$ color D4-branes are suspended between 
$NS5_L'$-brane and $NS5_R'$-brane. Moreover the $N_f$ D6-branes 
are located between the $NS5_L'$-brane and the NS5-brane 
and its mirrors $N_f$ D6-branes  
are located between the NS5-brane and the $NS5_R'$-brane.
The symmetric and conjugate symmetric flavors $S$ and  $\widetilde{S}$
are 4-4 strings stretching between D4-branes
located at the left hand side of O6-plane and those at the right hand
side of O6-plane  and furthermore
$N_f$ fundamental flavors $Q$ and $\widetilde{Q}$  are strings
stretching between $N_f$ D6-branes and $N_c$ color D4-branes. 

Let us deform this theory which has vanishing superpotential
by 1) adding the mass term for fundamental quarks 
and 2) the quartic term for fundamental quarks \footnote{The
deformation in \cite{Ahn07} is different from the quartic term
deformation of quarks since we do not rotate the NS-branes 
in this paper. That is,
after we take infinite mass limit for the adjoint field or equivalently
once the outer NS-branes become NS5'-branes, then 
one considers the above two
deformations 1) and 2).}.
The former can be achieved by ``displacing'' the D6-branes along $\pm v$
direction leading to their coordinates $v = \pm v_{D6}$ \cite{GK98} 
while the latter can be obtained by ``rotating'' the D6-branes
\cite{GK0710-1} 
by an angle 
$\theta$ in $(w,v)$-plane. Let us denote them by $D6_{\theta}$-branes
which are at angle $\theta$ with undeformed unrotated D6-branes(0123789).
Then their mirrors $N_f$ D6-branes are 
rotated by an angle $-\theta$ in $(w,v)$-plane according to O6-plane action and 
one denotes them by $D6_{-\theta}$-branes \footnote{The convention for
$D6_{\theta}$-branes here
is different from the one in \cite{Ahn07} where the angle between
unrotated D6-branes and $D6_{\theta}$-branes was not $\theta$ but 
$(\frac{\pi}{2}-\theta)$.}. 
Then, in the electric gauge theory, the most general 
deformed superpotential is
\bea
W_{elec} & = & \frac{\alpha}{2} \tr (Q \widetilde{Q})^2 - m \tr Q
\widetilde{Q} -\frac{1}{2\mu} \left[ (S\widetilde{S})^2 + Q
  \widetilde{S} S \widetilde{Q} +(Q\widetilde{Q})^2 \right] , \nonu \\
&& \mbox{with} \qquad \alpha = \frac{\tan \theta}{\Lambda}, \qquad 
m = \frac{v_{D6}}{2\pi \ell_s^2}
\label{electricsuperpotential}
\eea 
where $\Lambda$ is related to the scales of the electric and magnetic 
theories and $\pm v_{D6}$ is the $v$ coordinate of $D6_{\mp \theta}$-branes.
Of course, by rotating $D6_{\pm \theta}$-branes by $\mp \theta$
backwardly, 
giving rise
to unrotated $D6$-branes, 
and moving them to the origin of $v$, one obtains previous undeformed 
theory with vanishing superpotential.  
 Here the adjoint mass 
$\mu \equiv \tan \omega$
is related to a rotation angle $\omega$ 
of $NS5_{L,R}$-branes in $(v,w)$-plane. In the limit of $\mu
\rightarrow \infty$(or $\frac{\pi}{2}$ rotation of $NS5_{L,R}$-branes 
which become $NS5_{L,R}'$-branes after rotation), the terms of
$\frac{1}{\mu}$ in (\ref{electricsuperpotential}) vanish.

Let us summarize the ${\cal N}=1$ supersymmetric electric brane
configuration with nonvanishing superpotential 
(\ref{electricsuperpotential}) 
in type IIA string theory as follows and draw this in
Figure 1:

$\bullet$ One NS5-brane in (012345) directions with $w=0=x^6$

$\bullet$ Two NS5'-branes in (012389) directions with $v=0$

$\bullet$ $N_c$ color D4-branes in (01236) directions with $v=0=w$

$\bullet$ $N_f$ $D6_{\pm \theta}$-branes in (01237)
directions and
two other directions in $(v,w)$-plane 

$\bullet$ O6-plane in (0123789) directions with $x^6=0=v$

This brane configuration can be obtained from the brane configuration
of \cite{GK0710-1} by adding O6-plane with appropriate mirrors as we
mentioned in the previous section. 

%%%%%%%%%%%%%%%%%%%%%%%%%%%%%%%%%%%%%%%%%%%%%%%%%%%%%%%%%
%%%%%%%%%%%%%%%%%%%%%%%%%%%%%%%%%%%%%%%%%%%%%%%%%%%%%%%%%
\begin{figure}[ht]
   \epsfxsize=4in 
\centerline{\epsffile{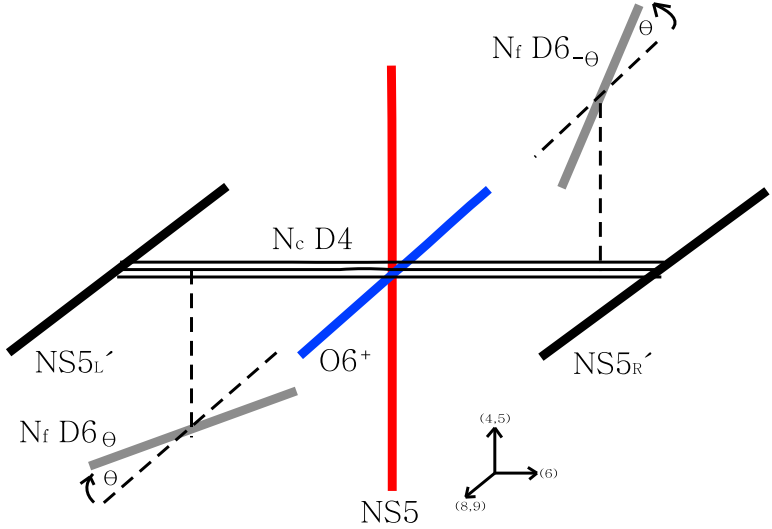}}
   \caption[FIG. \arabic{figure}.]{ 
The ${\cal N}=1$ 
supersymmetric electric brane configuration with deformed
superpotential (\ref{electricsuperpotential}) for the $SU(N_c)$ gauge
theory  with 
a symmetric flavor $S$, a conjugate symmetric flavor $\widetilde{S}$
and 
$N_f$ fundamental massive flavors $Q, \widetilde{Q}$. 
The origin of
the coordinates $(x^6, v, w)$ is located at the intersection of
NS5-brane and O6-plane. It is clear to see that the two deformations
are characterized by both displacement and rotation for D6-branes.  }
\end{figure}
%%%%%%%%%%%%%%%%%%%%%%%%%%%%%%%%%%%%%%%%%%%%%%%%%%%%%%%%%
%%%%%%%%%%%%%%%%%%%%%%%%%%%%%%%%%%%%%%%%%%%%%%%%%%%%%%%%%

Other different brane configuration can be obtained 
by moving the $D6_{\pm \theta}$-branes from Figure 1 into the outside of 
$NS5_{L,R}'$-branes. Since there are $N_f$ flavor D4-branes
connecting $D6_{\pm \theta}$-branes and the $NS5_{L,R}'$-branes after
this movement, the gauge singlet field $N$ appears. 
As noticed in \cite{GK0710-1}, at energies much below the mass
of $N$ the two brane descriptions 
coincide with each other. 
One regards this new brane configuration as integrating the field
$N$ in from Figure 1 and the superpotential of this electric theory
consists of the interaction term between $N$ with electric quarks,
mass term  and linear term for $N$ \cite{GK0710-1}.
The classical supersymmetric vacua of this brane configuration are 
classified by the parameter $k$ where $k=0,1, \cdots, N_c$ and
unbroken gauge symmetry in the $k$-th configuration is given by
$SU(N_c-k)$. In other words, the $k$ D4-branes among $N_f$ D4-branes 
(stretched between $NS5_R'$-brane and  $D6_{-\theta}$-branes) are 
reconnecting with
those number of D4-branes stretched between the middle NS5-brane and 
$NS5_R'$-brane. Then those recombined resulting 
$k$ D4-branes are moving to $\pm v$
direction and the remaining $(N_c-k)$ D4-branes are stretching between 
the middle NS5-brane and $NS5_R'$-brane and $(N_f-k)$ D4-branes are 
stretched between the $NS5_R'$-brane and $D6_{-\theta}$-branes(and
their mirrors).

%%%%%%%%%%%%%%%%%%%%%%%%%%%%%%%%%%%%%%%%%%%%%%%%%%%%%%%%%%%%%%%%%%%%% 
%%%%%%%%%%%%%%%%%%%%%%%%%%%%%%%%%%%%%%%%%%%%%%%%%%%%%%%%%%%%%%%%%%%%%%
\section{The ${\cal N}=1$ supersymmetric magnetic brane configuration}
%%%%%%%%%%%%%%%%%%%%%%%%%%%%%%%%%%%%%%%%%%%%%%%%%%%%%%%%%%%%%%%%%%%%%%
%%%%%%%%%%%%%%%%%%%%%%%%%%%%%%%%%%%%%%%%%%%%%%%%%%%%%%%%%%%%%%%%%%%%%%

The magnetic theory can be obtained by interchanging
the $D6_{\pm \theta}$-branes and $NS5_{L,R}'$-branes
while preserving the linking number. 
After we move the left $D6_{\theta}$-branes to the right all 
the way(and their
mirrors, right $D6_{-\theta}$-branes to the left) past the middle
NS5-brane 
and
the right $NS5_R'$-brane, the linking number counting \cite{Ahn07} 
implies that we should add $N_f$ D4-branes, corresponding to the meson
$M(\equiv Q \widetilde{Q})$, to the left 
side of all the right $N_f$ $D6_{\theta}$-branes(and their mirrors). 
Note that when a D6-brane crosses the middle NS5-brane,
due to the parallelness of these, there was no creation of D4-branes.
In other words, 
when the $D6_{\pm \theta}$-branes approach the middle NS5-brane, one
should take $\theta =\frac{\pi}{2}$ limit(making 
$D6_{\pm \frac{\pi}{2}}$-branes to be parallel to
the middle NS5-brane) and then after they cross the
middle NS5-brane, they return to the original positions 
given by $D6_{\pm \theta}$-branes as follows \footnote{For general brane setup
with arbitrary rotation angles of $D6_{\pm \theta}$-branes and
$NS_{\pm \theta'}$-branes, there are also other meson fields:$M_1
\equiv Q \widetilde{S} S \widetilde{Q}, P \equiv Q \widetilde{S} Q$ 
and $\widetilde{P}
\equiv \widetilde{Q} S \widetilde{Q}$. They couple to the dual quarks
and symmetric flavors in the superpotential. As stressed in \cite{Ahn07},
the particular geometric constraint at the intersection between
$D6_{\pm \frac{\pi}{2}}$-branes and the middle NS5-brane rules out the
presence of these gauge singlets, $M_1, P$ and $\widetilde{P}$.
\label{footnotespecial}}:     
\bea
D6_{\pm \theta}-\mbox{branes} \qquad \rightarrow \qquad 
 D6_{\pm \frac{\pi}{2}}-\mbox{branes} \qquad \rightarrow \qquad
D6_{\pm \theta}-\mbox{branes}.
\nonu
\eea

Next, we move the left $NS5_L'$-brane to the right all the way past
O6-plane
(and its mirror, right $NS5_R'$-brane to the left), and then the linking number
counting \cite{Ahn07} implies that the dual number of colors was $(2N_f-N_c)$.
There was no creation of D4-branes when the NS5'-brane crosses an
O6-plane because they are parallel to each other.
Now we draw the magnetic brane configuration in Figure 2 where 
some of the flavor D4-branes are recombining with those of
$(2N_f-N_c)$ color D4-branes and  those combined resulting D4-branes are moved
into $\pm v$ direction.
What happens for nonzero $k$'s vacua is as follows:
One takes $k$ D4-branes from $N_f$ flavor D4-branes and reconnect them
to those from $(2N_f-N_c)$ color D4-branes in Figure 2 such that
the resulting branes are connecting from the $D6_{\pm \theta}$-branes to
the middle NS5-brane directly. Their coordinates will be $v=\pm v_{D6}$ in
order to minimize the energy.
This Figure 2 also can be obtained from the magnetic brane configuration of 
\cite{GK0710-1} by adding O6-plane with right presence of mirrors
under the O6-plane action.

Then the low energy dynamics is described by the Seiberg dual magnetic
theory with gauge group $SU(2N_f-N_c)$, $N_f$ flavors of fundamentals 
$q, \widetilde{q}$, a symmetric flavor $s$, a conjugate symmetric
flavor $\widetilde{s}$, gauge singlet $M$ which is magnetic dual of
the electric meson field $Q \widetilde{Q}$.
Then the superpotential including the interaction between the meson
field $M$ and dual matters is described by  \footnote{If we consider
the finite adjoint mass $\mu$ with no interaction between the adjoint
field and quarks  and we do not consider the deformations by $m$ and
$\alpha$, the electric and magnetic theory is exactly the same as
the one in \cite{ILS} where the dual gauge group is given by $SU(3N_f+4-N_c)$.
That is, the electric superpotential contains only a quartic term of
symmetric tensor.
As in footnote \ref{footnotespecial}, there exist four terms in the
dual superpotential: a quartic term in symmetric tensor, $M_1$-, $P$- and
$\widetilde{P}$-term. If we allow the interaction between 
the adjoint
field and quarks, then the electric superpotential contains $Q
\widetilde{S} S \widetilde{Q}$ also and this plays the role of  
$M_1$ in the magnetic theory. The linear terms in $M_1$($M_1$ term and
$M_1 q \widetilde{q}$ term) force the vacuum
expectation
values for the dual quarks. Since $M_1$ is a $N_f \times N_f$ matrix
in flavor indices, all dual quarks get vacuum expectation
values. They break the magnetic gauge group $SU(3N_f+4-N_c)$ to 
$SU(2N_f+4-N_c)$. Furthermore, perturbing the electric superpotential
with the operators $Q \widetilde{S} Q$ and $\widetilde{Q} S
\widetilde{Q}$
implies that we add $P$ and $\widetilde{P}$ in the magnetic
superpotential.
Via the equations of motion for $P$ and $\widetilde{P}$ this forces
that    $q \widetilde{s} q$ and $\widetilde{q} s
\widetilde{q}$ have nonzero vacuum expectation values. Then the dual gauge
group $SU(2N_f+4-N_c)$ 
is further broken by two units per $q$ flavor and two units per 
$\widetilde{q}$ flavor. Note that 
the O6-plane has four D6-brane charge. 
Then the remaining dual gauge group is given by
$SU(2N_f-N_c)$ as we found from the brane configuration above. 
See the
analysis for this kind of vacuum expectation values in \cite{PS}.}
\bea
W_{mag} = \frac{1}{\Lambda^3} M q  \widetilde{s} s \widetilde{q} +  
\frac{\alpha}{2} \tr M^2- m \tr M, \qquad 
M \equiv Q\widetilde{Q}, \qquad
\alpha = \frac{\tan \theta}{\Lambda}, \qquad 
m = \frac{v_{D6}}{2\pi \ell_s^2}
% +  \left[ (s \widetilde{s})^2+
%M_1 q \widetilde{q}+ P_0 q \widetilde{s} q +
%\widetilde{P}_0 \widetilde{q} s \widetilde{q} + \cdots \right] 
\label{mag}
\eea
where the last two terms originate from 
the two deformations (\ref{electricsuperpotential}) from electric theory. 
The $\theta$-dependent coefficient function, $\alpha$, in front of 
quadratic term of the meson field
also occurs in the geometric brane interpretation for the different
supersymmetric gauge theory \cite{Ahn06}. 
The $\alpha=0$ limit reduces to the theory 
given by \cite{Ahn07}.

%%%%%%%%%%%%%%%%%%%%%%%%%%%%%%%%%%%%%%%%%%%%%%%%%%%%%%%%
%%%%%%%%%%%%%%%%%%%%%%%%%%%%%%%%%%%%%%%%%%%%%%%%%%%%%%%%%
\begin{figure}[ht]
   \epsfxsize=4in 
\centerline{\epsffile{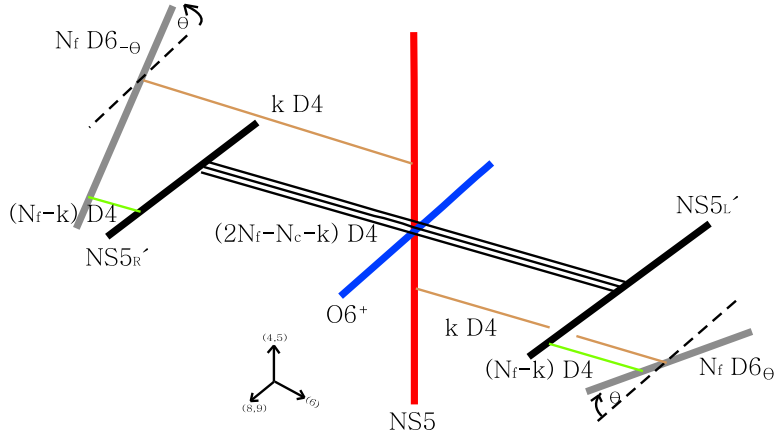}}
   \caption[FIG. \arabic{figure}.]{ 
The ${\cal N}=1$ supersymmetric magnetic 
brane configuration for the $SU(2N_f-N_c)$ gauge
theory  with 
a symmetric flavor $s$, a conjugate symmetric flavor $\widetilde{s}$ and 
$N_f$ fundamental flavors $q, \widetilde{q}$.
The $N_f$ flavor D4-branes connecting between
$NS5_L'$-brane and $D6_{\theta}$-branes are related to the dual gauge singlets
$M$ and are splitting into $(N_f-k)$ and
$k$ D4-branes. 
The location of intersection between $D6_{\theta}$-branes and $(N_f-k)$
D4-branes 
is given by $(v,w)=(0, v_{D6} \cot \theta)$ while the one between  
$D6_{\theta}$-branes and $k$
D4-branes 
is given by $(v,w)=(-v_{D6}, 0)$. The other end point of
 $(N_f-k)$
D4-branes has also $(v,w)=(0, v_{D6} \cot \theta)$ and they are finite
interval along $x^6$ direction.  }
\end{figure}
%%%%%%%%%%%%%%%%%%%%%%%%%%%%%%%%%%%%%%%%%%%%%%%%%%%%%%%%% 
%%%%%%%%%%%%%%%%%%%%%%%%%%%%%%%%%%%%%%%%%%%%%%%%%%%%%%%%%

In the magnetic theory, the superpotential is given by 
(\ref{mag}) and
in order to obtain the supersymmetric vacua, one solves the F-term equations:
\bea
\widetilde{s} s \widetilde{q} M & = & 0, \qquad s \widetilde{q} M q =0, 
\nonu \\
\widetilde{q} M q \widetilde{s} & = & 0, \qquad M q \widetilde{s}
s =0,
\nonu \\
\frac{1}{\Lambda^3} q \widetilde{s} s \widetilde{q} & = & m -\alpha M. 
\label{Fterm}
\eea
The last relation, by multiplying $M$ both sides,  
implies the following matrix equation
$
m M = \alpha M^2$.
Since the eigenvalues are either $0$ or $\frac{m}{\alpha}$, one takes
$N_f \times N_f$ matrix with $k$'s eigenvalues $0$ and $(N_f-k)$'s
eigenvalues $\frac{m}{\alpha}$:
\bea
M = \left(
\begin{array}{cc}
0 & 0  \\
0 & \frac{m}{\alpha} {\bf 1}_{N_f-k}
\end{array}
\right)
\label{M0}
\eea
where $k=1, 2, \cdots, N_f$ and ${\bf 1}_{N_f-k}$ is the $(N_f-k) \times
(N_f-k)$ identity matrix \cite{GK0710}.
The expectation value of $M$ can be represented by 
the fundamental string between the flavor brane displaced by the 
$w$ direction and the color brane from Figure 2. 
Therefore, in the brane configuration of Figure 2, the $k$ of the
$N_f$ flavor D4-branes are connected with $k$ of $(2N_f-N_c)$ color
D4-branes
and the resulting D4-branes stretch from the $D6_{\theta}$-branes to
the NS5-brane and the coordinate of an intersection point between the 
$k$ D4-branes and the NS5-brane is given by $(v, w)=(-v_{D6}, 0)$.
This corresponds to  exactly the $k$'s eigenvalues $0$ of $M$ above.
Now the remaining $(N_f-k)$ flavor D4-branes between 
the $D6_{\theta}$-branes and 
the $NS5_L'$-brane are related to the corresponding eigenvalues 
of $M$,  $\frac{m}{\alpha} {\bf 1}_{N_f-k}$.
The coordinate of an intersection point between the 
$(N_f-k)$ D4-branes and the $NS5_L'$-brane is given 
by $(v, w)=(0, v_{D6} \cot \theta)$ \footnote{Using the expressions
for $\alpha$ and $m$ from (\ref{mag}), one gets $\frac{m}{\alpha} =
\Lambda 
\frac{v_{D6} \cot \theta}{2\pi \ell_s^2}$ which should be 
$\Lambda \frac{w}{2\pi
\ell_s^2}$. 
Then $w=v_{D6} \cot \theta$.}.

After we substitute (\ref{M0}) into the last equation of (\ref{Fterm})
gives rise to 
\bea
q \widetilde{s} s \widetilde{q} = \left(
\begin{array}{cc}
m \Lambda^3 {\bf 1}_k & 0  \\
0 & 0
\end{array}
\right).
\label{solqq}
\eea
Since the rank of the left hand side of this is at most $2N_f-N_c$,
one must have more stringent bound $k \leq (2N_f-N_c)$.
In the $k$-th vacuum the gauge symmetry is broken to $SU(2N_f-N_c-k)$
and 
the supersymmetric vacuum drawn in Figure 2 with $k=0$ has 
$q \widetilde{s} = s \widetilde{q} =0$ and the gauge group 
$SU(2N_f-N_c)$ is unbroken. The expectation value of $M$ in this case 
is given by
$M = \frac{m}{\alpha} {\bf 1}_{N_f}= m \Lambda \cot \theta {\bf 1}_{N_f}$.

Other different brane configuration can be obtained 
by moving the D6-branes into the place between the middle NS5-brane
and the $NS5_{L,R}'$-branes. Since there are no flavor D4-branes
connecting $D6_{\pm \theta}$-branes and the $NS5_{L,R}'$-branes after
this movement, the meson field $M$ is gone. 
As noticed in \cite{GK0710-1}, at energies much below the mass
of $M$, one regards this brane configuration as integrating the field
$M$ out in Figure 2(corresponding to an integrating the field
$N$ in from Figure 1 of an electric gauge theory) 
and the superpotential of this magnetic theory
consists of the quadratic term  and quartic
term of magnetic dual quarks with nonzero vacuum expectation values of
$s$ and $\widetilde{s}$.

Another deformation arises when we rotate the $NS5_{L,R}'$-branes by
an angle $\pm \theta'$ in the $(v,w)$-plane. This rotation provides an
adjoint field of $SU(2N_f-N_c)$ and couples to the magnetic dual
quarks and the magnetic dual symmetric flavors. Integrating this
adjoint field out implies that there exists a further contribution 
to the quartic 
superpotential  for $q$ and $\widetilde{q}$ as well as the product of 
quadratic
term for $q, \widetilde{q}$ and quadratic
term for $s, \widetilde{s}$, and quartic term in $s$ and 
$\widetilde{s}$. In particular,
when the rotated $NS5_{\pm \theta'}$-branes are parallel to 
rotated $D6_{\pm \theta}$-branes, the coupling in front of 
$M^2$ in the magnetic
superpotential vanishes. See also the relevant paper \cite{HM}
described in different context(direct gauge mediation).

%%%%%%%%%%%%%%%%%%%%%%%%%%%%%%%%%%%%%%%%%%%%%%%%%%%%%%%%%%%%%%%
\section{Nonsupersymmetric meta-stable brane configuration }
%%%%%%%%%%%%%%%%%%%%%%%%%%%%%%%%%%%%%%%%%%%%%%%%%%%%%%%%%%%%%%%%

The theory has many nonsupersymmetric meta-stable ground states
besides the supersymmetric ones we discussed in previous section.
For the IR free region \cite{Ahn07}, $\frac{N_c}{2} < N_f < N_c +2$, 
the magnetic theory is the effective low energy description of the
asymptotically free electric gauge theory.
When we rescale the meson field as
$
M = h \Lambda \Phi $,
then the Kahler potential for $\Phi$ is canonical and the magnetic
quarks are canonical near the origin of field space.
The higher order corrections of Kahler potential are negligible when 
the expectation values of the fields $q, \widetilde{q}, s,
\widetilde{s}$ 
and $\Phi$
are smaller than the scale of magnetic theory.
Then the magnetic superpotential can be written in terms of $\Phi$ or $M$
\bea
W_{mag} = \frac{h}{\Lambda^2} \Phi  q  \widetilde{s} s \widetilde{q} 
 +  
\frac{h^2 \mu_{\phi}}{2} \tr \Phi^2- h \mu^2 \tr \Phi =
 \frac{1}{\Lambda^3} M q  \widetilde{s} s \widetilde{q} +  
\frac{\alpha}{2} \tr M^2- m \tr M.
\nonu
\eea
From this, one can read off the following quantities
\bea
\mu^2 = m \Lambda \left(= \frac{v_{D6}}{ g_s \ell_s^3} \right), 
\qquad \mu_{\phi} = \alpha \Lambda^2 \left(= \frac{\tan \theta }{g_s
    \ell_s} \right), 
\qquad M = h \Lambda \Phi.
\nonu
\eea

The classical supersymmetric vacua given by (\ref{M0}) and
(\ref{solqq})
can be described as 
\bea
 h \Phi
 = \left(
\begin{array}{cc}
0 & 0  \\
0 & \frac{\mu^2}{\mu_{\phi}} {\bf 1}_{N_f-k}
\end{array}
\right), \qquad
q \widetilde{s} s \widetilde{q} = \left(
\begin{array}{cc}
\mu^2 {\bf 1}_k & 0  \\
0 & 0
\end{array}
\right).
\nonu
\eea
Now one splits, as in \cite{GK0710-1,GK0710}, 
the $(N_f-k) \times (N_f-k)$
block  at the lower right corner of $h\Phi$ and $q \widetilde{s} s
\widetilde{q}$ 
into blocks of 
size $n$ and $(N_f-k-n)$ as follows:
\bea
h\Phi = \left(
\begin{array}{ccc}
0 & 0 & 0  \\
0 & h \Phi_n & 0 \\
0 & 0 & \frac{\mu^2}{\mu_{\phi}} {\bf 1}_{N_f-k-n}
\end{array}
\right), \qquad
q \widetilde{s} s \widetilde{q} = \left(
\begin{array}{ccc}
\mu^2 {\bf 1}_k & 0 & 0  \\
0 & { \varphi} \widetilde{\beta} \beta \widetilde{\varphi}  &  0 \\
0 & 0 & 0
\end{array}
\right).
\nonu
\eea
Here $\varphi$ and $\widetilde{\varphi}$ are $n \times (2N_f-N_c-k)$
dimensional matrices and correspond to $n$ flavors of fundamentals of
the gauge group $SU(2N_f-N_c-k)$ which is unbroken by the nonzero
expectation value of $q$ and $\widetilde{q}$.
In the brane configuration in Figure 3, 
they correspond to 
fundamental strings connecting the $n$ flavor D4-branes and
$(2N_f-N_c-k)$
color D4-branes.
This Figure 3 can be obtained from the meta-stable brane configuration of 
\cite{GK0710-1} by adding O6-plane with appropriate mirrors
under the O6-plane action.
The symmetric and conjugate symmetric flavors $\beta$ and  $\widetilde{\beta}$
are 4-4 strings stretching between $(2N_f-N_c-k)$ D4-branes
located at the left hand side of O6-plane and those at the right hand
side of O6-plane in Figure 3.
The $\Phi_n$ and ${ \varphi} \widetilde{\beta} \beta
\widetilde{\varphi}$
are $n \times n$ matrices.
The supersymmetric ground state corresponds to
$h\Phi_n= \frac{\mu^2}{\mu_{\phi}} {\bf 1}_{n}, 
\varphi =\widetilde{\beta}=\beta =\widetilde{\varphi}=0$. 

Now the full one loop potential for $\Phi_n, \hat{\varphi} \equiv \varphi \widetilde{\beta}, 
\hat{\widetilde{\varphi}} \equiv \beta \widetilde{\varphi} $
\cite{Ahn07} 
takes the form
\bea
\frac{V}{|h|^2}  =  
|\Phi_n  \hat{\varphi}|^2   
+  |\Phi_n  \hat{\widetilde{\varphi}}  |^2
  +  
| \hat{\varphi}  \hat{\widetilde{\varphi}}-\mu^2 {\bf 1}_{n} + 
h \mu_{\phi} \Phi_n|^2 + b |h \mu|^2 \tr \Phi_n^{\dagger} \Phi_n, 
\; b = \frac{(\ln 4-1)}{8\pi^2} (2N_f-N_c).
\nonu
\eea
Differentiating this potential with respect to 
$\Phi_n$ and putting $\hat{\varphi}=0, \hat{
\widetilde{\varphi}}=0$, one obtains
\bea
h \Phi_n = \frac{\mu^2 \mu_{\phi}^{\ast}}{|\mu_{\phi}|^2 +b |\mu|^2}
{\bf 1}_n
\simeq \frac{\mu^2 \mu_{\phi}^{\ast}}{b |\mu|^2}
{\bf 1}_n \qquad \mbox{or} \qquad
M_n \simeq \frac{\alpha \Lambda^3}{(2N_f-N_c)} {\bf 1}_{n}
\label{vac}
\eea
for real $\mu$ and 
we assume here that 
$\mu_{\phi} << \mu << \Lambda_m$. The vacuum energy $V$ is given by
$V \simeq n |h \mu^2|^2$.
Expanding around this solution, one obtains
the eigenvalues for mass matrix for $\hat{\varphi}$ and 
$\hat{\widetilde{\varphi}}$ 
will be
\bea
m_{\pm}^2 = \frac{|\mu|^4}{(|\mu_{\phi}|^2 + b |\mu|^2)^2}\left[ 
|\mu_{\phi}|^2 \pm b |h|^2 \left( |\mu_{\phi}|^2 + b|\mu|^2 \right)
\right]
\simeq \frac{1}{b^2} \left( |\mu_{\phi}|^2  \pm |b h \mu|^2 \right).
\nonu
\eea
Then  for 
$| \frac{\mu_{\phi}}{\mu}  |^2 
> \frac{|b h|^2}{1-b|h|^2} \simeq |bh|^2
$ in order to avoid tachyons
the vacuum (\ref{vac}) is locally stable.

Recall that the expectation value of $\Phi$ can be represented by 
the fundamental string between the flavor brane displaced by the 
$w$ direction and the color brane from Figure 3. 
One can move $n$ D4-branes, from $(N_f-k)$ D4-branes stretched
between the $NS5_L'$-brane and the $D6_{\theta}$-branes at $w=v_{D6}
\cot \theta $, to the local minimum of the potential and the end
points of these $n$ D4-branes are at a nonzero $w$ as in Figure 3.
Alternatively, in the brane configuration of Figure 3, the $k$ of the
$(N_f-n)$ flavor D4-branes are connected with $k$ of $(2N_f-N_c)$ color
D4-branes
and the resulting D4-branes stretch from the $D6_{\theta}$-branes to
the NS5-brane and the coordinate of an intersection point between the 
$k$ D4-branes and the NS5-brane is given by $(v, w)=(-v_{D6}, 0)$.
This corresponds to  exactly the $k$'s eigenvalues $0$ of $h \Phi$ above.
The remaining $(N_f-k-n)$ flavor D4-branes between 
the $D6_{\theta}$-branes and 
the $NS5_L'$-brane are related to the corresponding eigenvalues 
of $h\Phi$,   $\frac{\mu^2}{\mu_{\phi}} {\bf 1}_{N_f-k-n}$.
The coordinate of an intersection point between the 
$(N_f-k-n)$ D4-branes and the $NS5_L'$-brane is given 
by $(v, w)=(0, v_{D6} \cot \theta)$.
Finally, 
the remnant $n$ flavor ``curved'' D4-branes between 
the $D6_{\theta}$-branes and 
the $NS5_L'$-brane are related to the corresponding eigenvalues 
of $h\Phi_n$ by (\ref{vac}).
Note that since the eigenvalues of $\Phi$ characterizes the $w$
coordinate of flavor D4-branes and $  \frac{\mu^2 
\mu_{\phi}^{\ast}}{b |\mu|^2} << \frac{\mu^2}{\mu_{\phi}}$,
the $n$ D4-branes are nearer to the $w=0$ located at the NS5-brane.

As pointed out in \cite{GK0710-1}, this local stable vacuum decays to
the supersymmetric ground states. The end points of $n$ ``curved'' flavor
D4-branes
on the $NS5_L'$-brane approach those  of the $(2N_f-N_c-k)$ color 
D4-branes and two types of branes reconnect each other. 
For $n \leq (2N_f-N_c-k)$, the final brane configuration is nothing
but the supersymmetric vacuum of Figure 2 with the replacement 
$k \rightarrow (k+n)$.
When $n > (2N_f-N_c-k)$, then the remnant $[n-(2N_f-N_c-k)]$ 
flavor D4-branes remain.   
On the other hand, the $n$ D4-branes can move to larger $w$ and return
to the Figure 2. Also some of the D4-branes approach the intersection
point between $D6_{\theta}$-branes and the $NS5_L'$-brane while the
remaining D4-branes move to the one between $D6_{\theta}$-branes and
the NS5-brane.  

%%%%%%%%%%%%%%%%%%%%%%%%%%%%%%%%%%%%%%%%%%%%%%%%%%%%%%%%%%
%%%%%%%%%%%%%%%%%%%%%%%%%%%%%%%%%%%%%%%%%%%%%%%%%%%%%%%%%
\begin{figure}[ht]
   \epsfxsize=4in 
\centerline{\epsffile{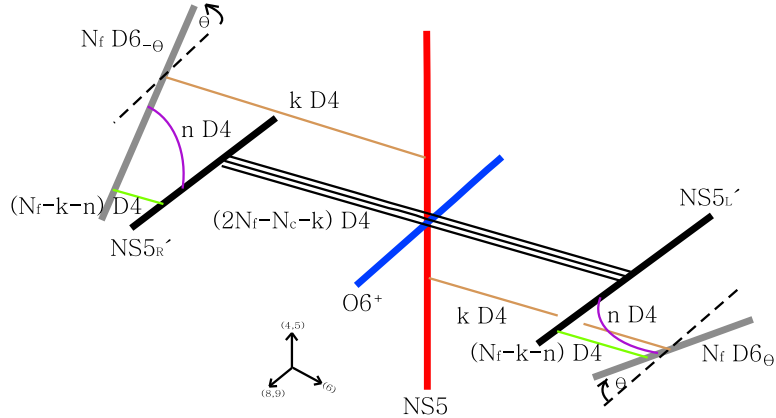}}
   \caption[FIG. \arabic{figure}.]{ 
The nonsupersymmetric minimal energy
brane configuration for the $SU(2N_f-N_c)$ gauge theory  with 
a symmetric flavor $s$, a conjugate symmetric flavor $\widetilde{s}$ and 
$N_f$ fundamental flavors $q, \widetilde{q}$.
This brane configuration can be obtained by moving $n$ flavor D4-branes, 
from $(N_f-k)$ flavor D4-branes stretched
between the $NS5_L'$-brane and the $D6_{\theta}$-branes in Figure
2(and their mirrors). The $w$ coordinate for $n$ flavor ``curved'' D4-branes is
determined and is given by (\ref{relation}) or (\ref{wvalue}). }
\end{figure}
%%%%%%%%%%%%%%%%%%%%%%%%%%%%%%%%%%%%%%%%%%%%%%%%%%%%%%%%%
%%%%%%%%%%%%%%%%%%%%%%%%%%%%%%%%%%%%%%%%%%%%%%%%%%%%%%%%%

In the remaining paragraphs, 
we focus on the gravitational potential of the NS5-brane.
The essential features of 
all these considerations already appeared in \cite{GK0710-1}.
Let us remind that the branes are placed as follows:
\bea
D6_{\theta}-\mbox{branes}(01237vw) & : & \qquad v  =  -v_{D6} +  w \tan
\theta, \nonu \\
NS5'_L-\mbox{brane}(012389) & : & \qquad  v  =  0, \nonu \\
NS5-\mbox{brane}(012345) & : & \qquad w  =  0
\nonu
\eea
where we assume that $D6_{\theta}$-branes and $NS5_L'$-brane 
are located at the same value of 
\bea
y
\equiv x^6.
\nonu 
\eea 
The generalization for different $y$'s can be done similarly. 
The dependence of the distance between the $D6_{\theta}$-branes and 
$NS5_L'$ brane along the NS5-brane on $w$ can be represented by
\bea
\Delta x =|-v_{D6} +  w \tan \theta |.
\label{wdependence}
\eea
Due to this fact, the partial differentiation of $\Delta x$ with
respect to
$w$ will lead to an extra contribution for the computation of 
$\pa_w \theta_w$ where we define $\theta_w$ as 
\bea
\theta_w \equiv
\cos^{-1} \left(\frac{y_m}{\sqrt{y^2+|w|^2}}
\right)
\label{thetaw}
\eea
with $y_m$ that is smallest value of $y$ along the D4-brane. 
Then the energy density of the D4-brane was computed in 
\cite{GK0703,Ahn07-5}
and is given by
\bea
E(w) = 2 \tau_4 \frac{y_m \sqrt{y^2+|w|^2}}{\ell_s} \sqrt{1+
\frac{\ell_s^2}{y_m^2}} \sin \theta_w
\label{energy}
\eea
corresponding to (3.7) of \cite{GK0710-1} 
where $\tau_4$ is a tension of D4-brane in flat spacetime.

It is straightforward to compute 
the differentiation of $\left( \frac{\ell_s}{2 \tau_4} E(w)\right)^2$
with respect to $w$ and it leads to
\bea
\pa_w \left( \frac{\ell_s}{2 \tau_4} E(w)\right)^2  & = &
\ell_s^2 \bar{w} \sin^2 \theta_w +\ell_s^2 (y^2+ |w|^2) \sin 2\theta_w
\pa_w \theta_w + \frac{1}{2} (y^2+|w|^2) \bar{w} \sin^2 2\theta_w
\nonu \\
& + &
\frac{1}{2}  (y^2+|w|^2)^2 \sin 4\theta_w \pa_w \theta_w.
\label{equ}
\eea
In order to simplify this, one uses the partial differentiation of
$\Delta x$ (\ref{wdependence}) with respect to $w$ which is 
equal to $\frac{1}{2} 
\tan \theta  \frac{\bar{w} \bar{\tan \theta}  -\bar{v_{D6}}}{|w \tan
\theta  -v_{D6}|}$. On the other hand, $\Delta x$ was known in
\cite{GK0703}
and it is $\frac{(y^2+|w|^2)}{\ell_s} \sin 2\theta_w + 2\ell_s \theta_w$.
After differentiating this with respect to $w$ and equating it to the
previous expression, one arrives at 
\bea
[\ell_s^2 + (y^2 +|w|^2) \cos 2\theta_w ] \pa_w \theta_w  +\frac{1}{2}
\bar{w} \sin 2\theta_w -\frac{1}{2}\ell_s \frac{1}{2} 
\tan \theta  \frac{\bar{w} \bar{\tan \theta}  -\bar{v_{D6}}}{|w \tan
\theta  -v_{D6}|} =0
\nonu
\eea
corresponding to (3.21) of \cite{GK0710-1}
and putting this into 
(\ref{equ}) one gets 
\bea
\pa_w \left( \frac{\ell_s}{2 \tau_4} E(w)\right)^2 =
\ell_s^2 \bar{w} \sin^2 \theta_w +\frac{1}{4} \ell_s \tan \theta 
(y^2 +|w|^2) \sin 2\theta_w  \frac{\bar{w} \bar{\tan \theta}
-\bar{v_{D6}}}{|w \tan
\theta  -v_{D6}|}.
\label{rel}
\eea
It is easy to see that at $w = v_{D6} \cot \theta$ which is an
intersection point between $D6_{\theta}$-branes and $NS5_L'$-brane, 
$\Delta x$ vanishes through (\ref{wdependence}) and this also implies 
that $\theta_w$ vanishes. Furthermore, 
the energy 
$E(w)$ is zero from (\ref{energy}). This corresponds to the global
minimal energy. For the parallel $D6$-branes and $NS5_L'$-brane(i.e.,
$\tan \theta =0$), then the only stationary point is $w=0$
\cite{Ahn07}. 
If $w \neq
0$, then $\theta_w =0$ or $\theta_w =\frac{\pi}{2}$ 
from the first term of (\ref{rel}) but these are not physical solutions. 
 
For real and positive parameters $v_{D6}, w$ and $\tan \theta$, we are
looking for the solution with $v_{D6} > w \tan \theta $
and setting the right hand side of (\ref{rel}) to zero,
finally one gets with (\ref{thetaw})
\bea
\tan \theta_w = \frac{ (y^2 + w^2) \tan \theta }{2 w \ell_s}.
\label{relation}
\eea
Therefore, the brane configuration of Figure 3 has a local minimum
where the end of D4-brane are located at $w$ given by (\ref{relation}).
When the $\theta_w$ is small, one can approximate 
(\ref{relation}) and substitute it into the $\Delta x$, one gets
\bea
w \simeq \tan \theta \frac{y^4}{\ell_s^2 v_{D6}}.
\label{wvalue}
\eea
According to the analysis of \cite{GK0710-1}, 
the gauge theory result is valid only when $\theta$ and
$\frac{v_{D6}}{\ell_s}$
are much smaller than $g_s$.
On the other hand, the classical brane construction with
(\ref{relation})
is valid for any angle and the length parameters are of order $\ell_s$
or larger. 

%%%%%%%%%%%%%%%%%%%%%%%%%%%%%%%%%%%%%%%%%%%%%%%%%%%%%%%%%%%%%%%
%%%%%%%%%%%%%%%%%%%%%%%%%%%%%%%%%%%%%%%%%%%%%%%%%%%%%%%%%%%%%%%
\section{Conclusions and outlook}
%%%%%%%%%%%%%%%%%%%%%%%%%%%%%%%%%%%%%%%%%%%%%%%%%%%%%%%%%%%%%%%%
%%%%%%%%%%%%%%%%%%%%%%%%%%%%%%%%%%%%%%%%%%%%%%%%%%%%%%%%%%%%%%%%

We have constructed the type IIA brane configuration, presented in
Figure 3,
corresponding to 
the meta-stable supersymmetry breaking vacua for 
${\cal N}=1$ $SU(N_c)$ supersymmetric gauge
theory with a symmetric flavor, a conjugate symmetric flavor and
fundamental flavors when there exists  
a quartic term in the superpotential as well as the mass term for quarks. 
The gravitational attraction of D4-branes to the NS5-brane 
leads to the meta-stable states and this feature is similar to the
case for the gauge
theory meta-stable ground states.
Basically the meta-stable brane configuration we present is obtained from 
\cite{GK0710-1} by adding O6-plane with correct mirror branes.

It is natural to ask whether the method of present paper or the work
of \cite{GK0710}
can apply to other supersymmetric gauge theories which can be realized
in terms of type IIA string theory.
There exist many SQCD-like theories \cite{ILS} where the
Seiberg dual theories are known explicitly. 
For example, it would be interesting to deform the theories given in 
\cite{Ahn07-4,Ahn07-5,Ahn07-7,Ahn07-9,Ahn07-10} where one of the gauge
group factor has the same matter contents as the one of the present paper 
and see how the meta-stable ground states appear in the gauge theory
analysis or string theory description.
Or it is an open problem to see
the application for different O6-plane charge studied in \cite{Ahn07-1,Ahn07-3}.

When an adjoint matter field is included in the gauge theory, then we
need to increase the number of NS5-branes \cite{GK98} from the brane
configuration.
Also it is possible to 
deform the superpotential by quartic term for the quarks into the
description of \cite{Ahn06}. 
Along the lines of 
\cite{GK0703,Ahn07-5,Ahn07-6,Ahn07-7,Ahn07-10}, when the
$D6_{\theta}$-branes
are replaced by a single $NS5_{\theta}$-brane, this procedure is related to the
gauging the global flavor symmetry. It would be interesting to see how
the present deformation arises in these theories explicitly.  
As suggested in \cite{GK0710-1}, it would be interesting to see what
happens if the $NS5_{L,R}'$-branes are replaced by other D6-branes.
It is also possible to deform the symplectic or orthogonal gauge
group theory with massive flavors 
\cite{Ahn06-1} by adding an orientifold 4-plane to the brane
configuration of \cite{GK0710-1}. In the gauge theory side, 
one adds the quartic deformation term for the quarks 
in the superpotential. Similar application to the product gauge group
case \cite{Ahn07-2,Ahn07-8} is also possible.

\vspace{.7cm}

%%%%%%%%%%%%%%%%%%%%%%%%%%%%%%%%%%
\centerline{\bf Acknowledgments}
%%%%%%%%%%%%%%%%%%%%%%%%%%%%%%%%%%

We would like to thank D. Kutasov for discussions.
This work was supported by grant No.
R01-2006-000-10965-0 from the Basic Research Program of the Korea
Science \& Engineering Foundation.

\end{document}